\documentclass[pra,aps,twocolumn,superscriptaddress,notitlepage]{revtex4-1}
\usepackage{amssymb,amsfonts,amsmath,amstext,graphicx,hyperref}
\usepackage{tikz}
\usepackage{graphicx}
\usepackage[normalem]{ulem}
\hypersetup{
	colorlinks=true,
	linkcolor=blue,
	filecolor=magenta,      
	urlcolor=cyan,
}
\usepackage{xcolor}

\begin{document}

\title{Seeding crystallization in time}

\author{Michal Hajdu\v{s}ek}
\email{michal@sfc.wide.ad.jp}
\affiliation{Keio University Shonan Fujisawa Campus, 5322 Endo, Fujisawa, Kanagawa 252-0882, Japan}
\affiliation{Keio University Quantum Computing Center, 3-14-1 Hiyoshi, Kohoku, Yokohama, Kanagawa 223-8522, Japan}

\author{Parvinder Solanki}
\affiliation{Department of Physics, Indian Institute of Technology-Bombay, Powai, Mumbai 400076, India}

\author{Rosario Fazio}
\affiliation{The Abdus Salam International Center for Theoretical Physics (ICTP), Strada Costiera 11, 34151 Trieste, Italy}
\affiliation{Dipartimento di Fisica, Universit\`{a} di Napoli Federico II, Monte S. Angelo, I-80126 Napoli, Italy}

\author{Sai Vinjanampathy}
\email{sai@phy.iitb.ac.in}
\affiliation{Department of Physics, Indian Institute of Technology-Bombay, Powai, Mumbai 400076, India}
\affiliation{Centre for Quantum Technologies, National University of Singapore, 3 Science Drive 2, 117543 Singapore, Singapore}

\date{\today}

\begin{abstract}
We introduce the concept of seeding of crystallization in time by studying the dynamics of an ensemble of coupled continuous time crystals.
We demonstrate that a single subsystem in a broken-symmetry phase acting as a nucleation center may induce time-translation symmetry breaking across the entire ensemble.
Seeding is observed for both coherent as well as dissipative coupling, and for a broad range of parameter regimes. In the spirit of mutual synchronization, we investigate the parameter regime where all subsystems are in the broken symmetry phase. We observe that more broadly detuned time crystals require weaker coupling strength to be synchronized. This is in contrast to basic knowledge from classical as well as quantum synchronization theory. We show that this surprising observation is a direct consequence of the seeding effect.

\end{abstract} 
\maketitle

\textit{Introduction.---}
Time crystals are non-equilibrium phases of matter with broken time-translation symmetry \cite{sacha2017time,else2020discrete}.
Initial theoretical investigations \cite{sacha2015modeling,else2016floquet,khemani2016phase,russomanno2017floquet,surace2017floquet,heugel2019classical,khasseh2019many,sakurai2021chimera} and experimental demonstrations so far \cite{pal2018temporal,rovny2018observation,smits2018observation,zhang2017observation,choi2017observation,kyprianidis2021observation,taheri2021all} focused on breaking of the \emph{discrete} time-translation symmetry in closed systems, followed soon after by dissipative systems as well \cite{gong2018discrete,lazarides2020time,campeny2020time,kessler2021observation,sakurai2021dephasing}.
Introducing dissipation led to the \emph{continuous time crystals} (CTCs) \cite{iemini2018boundary, tucker2018shattered,buca2019non,zhu2019dicke,lledo2019driven,seibold2020dissipative,prazeres2021boundary,piccitto2021symmetries,carollo2021exact}.
Here, the continuous time-translation symmetry is broken in the thermodynamic limit when a self-organized steady state emerges with period of oscillation depending only on the system parameters.
Emergence of global oscillations in coupled classical systems has been extensively investigated, particularly in chemistry \cite{tinsley2009emergence,taylor2009dynamical} and biology \cite{aldridge1976cell,dano1999sustained} in the context of quorum sensing \cite{singh2012alternate}.
Inspired by the growth of broken symmetry phases in classical phase transitions by seeding a crystal in a solution \cite{mullin2001crystallization} we investigate the following question.
Can the broken time-translation symmetry of a seeding crystal induce time crystallization of the entire system in a similar fashion?
We answer this question in the affirmative and demonstrate that seeding of crystallization in time is indeed not only possible but under certain conditions inevitable.

We investigate a network of $n$ interacting CTCs where all but one subsystem are in the unbroken-symmetry phase.
We derive the required conditions under which the time-translation symmetry breaking permeates across the entire system.
We consider the thermodynamic limit and also demonstrate that this effect occurs for finite time-scales in the finite-size scenario.
We analyze the eigenspectrum of the generator of dynamics to understand the mechanism of crystallization of time.
Furthermore, in the spirit of mutual synchronization \cite{pikovsky2003synchronization}, we consider the dynamics of the subsystems in the broken symmetry phase but oscillating at different natural frequencies.
We obtain the phase diagram of when synchronization of the two subsystems occurs and find that farther detuned oscillators can be more easily synchronized.
This is in stark contrast to current understanding of synchronization in both classical as well as quantum systems.
We show that this surprising behavior is a direct consequence of the seeding effect. In particular, we show that weakly coupled subsystems oscillate with distinct observed frequencies but upon reaching a critical coupling one of the subsystems begins to seed oscillations from the other leading to synchronized dynamics.

\begin{figure*}[t]
    \centering
    \includegraphics[width=\textwidth]{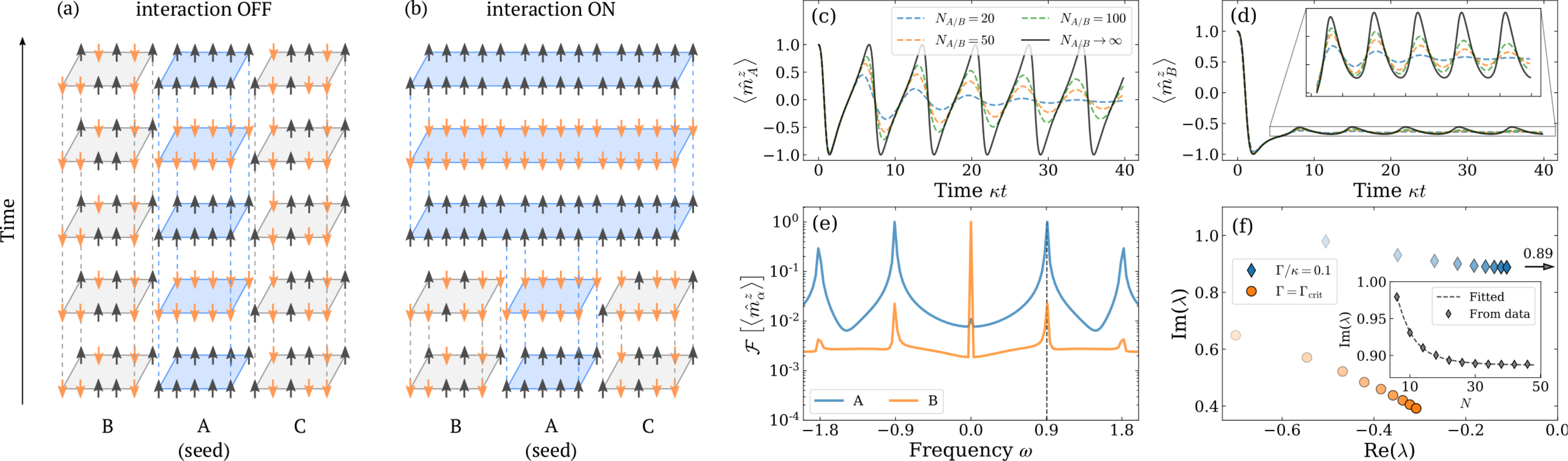}
    \caption{(a) In the absence of interactions crystallization (blue regions) does not permeate through the whole system. Subsystems B and C remain time-translationally symmetric. (b) When interaction is turned on, ensemble A seeds crystallization to ensembles B and C after some initial transient. (c)-(d) Time evolution of $\langle \hat{m}^z_{\alpha} \rangle$ for two subsystems A (seed) and B for finite size and the thermodynamic limit. Parameters are $\Omega_A/\kappa=1.5$, $\Omega_B/\kappa=0.9$ and $\Gamma/\kappa=0.1$. (e) Normalized Fourier transform of $\langle\hat{m}^z_{\alpha}\rangle$. The coinciding peaks show that $B$ is phase-locked to $A$. (f) Scaling of the dominant eigenvalue of the Liouville eigenspectrum for finite-sized subsystems when $N=\{6,10,14,18,22,26,30,34,38\}$. Increasing color saturation of the points represents increasing system size. Blue diamonds represent coupling of $\Gamma/\kappa=0.1 < \Gamma_{\text{crit}} / \kappa$ and seeding occurs. The inset shows functional scaling of $\text{Im}(\lambda)$ and sits convergence to $0.89$. The orange circles represent the case of critical coupling $\Gamma_{\text{crit}}$ when both subsystems are in the unbroken-symmetry phase. The real part of the dominant eigenvalue does not vanish anymore.}
    \label{fig:figure1}
\end{figure*}

\textit{Seeding Crystallization.---}
Before exploring a concrete example, we establish the general signatures of seeding of crystallization in time both in the thermodynamic limit as well as in the finite-size case.
Spontaneous symmetry breaking in a CTC is witnessed by the emergence of transient oscillations of an order parameter $\langle\hat{O}\rangle$ for finite size.
These oscillations become persistent in the thermodynamic limit \cite{iemini2018boundary} when the system size diverges.
Development of such persistent oscillations in generic driven dissipative systems is not guaranteed and depends on the interplay of coherent and dissipative dynamics, characterized by parameters $\Omega$ and $\kappa$, respectively.
Dynamics of such a system is governed by a Liouvillian superoperator,
\begin{equation}
    \dot{\rho} = L_{\text{CTC}}[\Omega,\kappa](\rho).
\end{equation}
Specific examples of such models are considered below. In a Dicke model of the CTC \cite{iemini2018boundary} crystallization occurs in the weakly dissipative regime, $\Omega/\kappa > 1$, while the system remains time-translationally invariant in the strongly dissipative regime, $\Omega/\kappa \leq 1$.

Consider a network of $n$ coupled CTCs evolving according to
\begin{equation}
    \dot{\rho} = \sum_{\alpha}L_{\text{CTC}}[\Omega_{\alpha},\kappa_{\alpha}](\rho) + L_V[\Gamma](\rho),
    \label{eq:me_BTC_many}
\end{equation}
where $\alpha$ indexes the subsystem and $L_V[\Gamma](\rho)$ describes their interaction at strength $\Gamma$.
All subsystems are in the strongly dissipative regime except for the seed which is set to be in the broken-symmetry phase, $\Omega_{\text{seed}}/\kappa_{\text{seed}} > 1$.
In the thermodynamic limit, seeding is witnessed by oscillating order parameters $\langle\hat{O}_{\alpha}\rangle$ signifying a broken-symmetry phase for any subsystem $\alpha$.
Fig.~\ref{fig:figure1}(a)-(b) illustrates this idea with an example of three subsystems where $A$ is the seed.

For finite subsystems, the time-translation symmetry is broken for finite time scales before the order parameter oscillations vanish (even for the seed).
Suitable tool for analyzing this scenario is the spectrum of the Liouville superoperator, denoted by complex eigenvalues $\lambda_j$, generating the dynamics of the system \cite{albert2014symmetriews,iemini2018boundary,campeny2020time,tindall2020quantum,solanki2021role,buca2021algebraic}.
The transient oscillations of the order parameter are a consequence of finite imaginary part of the eigenvalues, $\text{Im}(\lambda_j) \neq 0$, while their damping is the result of negative real parts, $\text{Re}(\lambda_j) < 0$.
Witnessing seeding in this case amounts to requiring the imaginary part of the Liouville eigenspectrum to remain gapped while its real part becomes gapless as the subsystem's size increases.

\textit{Model.---}
Consider $n$ subsystems indexed by $\alpha$, each composed of $N_{\alpha}$ spin-$1/2$ atoms with total spin $S_{\alpha}=N_{\alpha}/2$.
The spins undergo collective dissipation at rates $\kappa_{\alpha}/S_{\alpha}$ and are pumped coherently with strength $\Omega_{\alpha}$.
We now make a particular choice for the interaction $L_V[\Gamma]$ in Eq.~(\ref{eq:me_BTC_many}) and discuss other interactions later.
We assume the subsystems interact dissipatively at rate $\Gamma/S$, where $S=\sum_{\alpha}S_{\alpha}$.
Evolution of the total system is described by the following master equation (see Appendix \ref{sec:appendix-A}),
\begin{equation}
    \dot{\rho} = \sum_{\alpha} \left\{-i [\Omega_{\alpha} \hat{S}_{\alpha}^x, \rho] + \frac{\kappa_{\alpha}}{S_{\alpha}} \mathcal{D}[\hat{S}_{\alpha}^-]\rho \right\} + \frac{\Gamma}{S}\mathcal{D}[\hat{S}^-]\rho,
    \label{eq:master_eq}
\end{equation}
where $\hat{S}_{\alpha}^k=\frac{1}{2}\sum_j\hat{\sigma}_{\alpha,j}^k$ are the collective spin operators for $k=x,y,z$ and subsystem $\alpha$, $\hat{S}_{\alpha}^- = \hat{S}_{\alpha}^x - i \hat{S}_{\alpha}^y$ and $\hat{S}^- = \sum_{\alpha}\hat{S}_{\alpha}^-$ are the collective annihilation operators.
The dissipators are given in Lindblad form as $\mathcal{D}[\hat{O}]\rho = \hat{O}\rho\hat{O}^{\dagger} - \frac{1}{2} \{\hat{O}^{\dagger}\hat{O},\rho\}$.
For simplicity, we consider the subsystems with $\kappa_{\alpha} = \kappa$, and $S_{\alpha} = S_{\beta}$.

We begin with an analysis of Eq.~(\ref{eq:master_eq}) in the thermodynamic limit.
Note that the rescaled spin operators $\hat{m}_{\alpha}^k\equiv\hat{S}_{\alpha}^k/S_{\alpha}$, obey the commutation relation $[\hat{m}_{\alpha}^k, \hat{m}_{\alpha}^l] = i\varepsilon^{klm}\hat{m}_{\alpha}^m/S_{\alpha}$, where $\varepsilon^{klm}$ is the Levi-Civita symbol.
In the thermodynamic limit when $S_{\alpha} \rightarrow\infty$, these operators commute and their two-point correlation functions factorize \cite{iemini2018boundary}, leading to a system of classical nonlinear dynamical equations for the expectation values $\langle\hat{m}_{\alpha}^k\rangle$ (see the Appendix \ref{sec:appendix-B}).

Consider the case of $n=2$ subsystems labelled by $A$ and $B$, respectively.
Subsystem $A$ is the seed with $\Omega_A/\kappa > 1$ while subsystem $B$ is set in the strongly dissipative regime with $\Omega_B/\kappa < 1$.
In the absence of coupling, only $A$ breaks the time-translation symmetry.
Introducing weak coupling induces symmetry breaking in subsystem $B$ as depicted in Fig.~\ref{fig:figure1}(c)-(d) where we plot $\langle\hat{m}_A^z\rangle$ and $\langle\hat{m}_B^z\rangle$, respectively.
In the finite-size case, oscillations of $A$ and $B$ are transient but become persistent limit-cycle oscillations when $N_{A/B}\rightarrow\infty$ as expected from \cite{iemini2018boundary}.
Furthermore, we observe the induced oscillations in ensemble $B$ are phase-locked to the oscillations in $A$ by examining the Fourier transform $\mathcal{F}\left[\langle\hat{m}^z_{\alpha}\rangle\right]$ in Fig.~\ref{fig:figure1}(e).

It is important to note the role of dissipation in breaking the time-translation symmetry.
In the case of a single subsystem considered in \cite{iemini2018boundary}, the limit-cycle oscillations are a result of the interplay between coherent driving producing oscillations and collective dissipation, causing the damping of these oscillations.
Stable limit-cycles emerge only once the coherent drive overcomes the dissipation and $\Omega_{\alpha}/\kappa > 1$.
Therefore it is surprising that limit-cycle oscillations in Fig.~\ref{fig:figure1}(d) are seeded via dissipative coupling.
Furthermore, seeding occurs for arbitrarily small $\Gamma/\kappa$, albeit producing only arbitrarily small oscillations.

Increasing the coupling strength also increases the amplitude of seeded oscillations provided the coupling is below a certain critical strength, $\Gamma < \Gamma_{\text{crit}}$.
Beyond this critical coupling both subsystems enter the unboken-symmetry phase.
This critical coupling $\Gamma_{\text{crit}}$ can be obtained via linear stability analysis (see Appendix \ref{sec:appendix-B}),
\begin{equation}
    \Gamma_{\text{crit}} = \frac{2\kappa (\Omega_A - \kappa)}{2\kappa - (\Omega_A - \Omega_B)},
    \label{eq:Gamma_crit}
\end{equation}
where $\Omega_A > \Omega_B$. We note that increasing the coherent driving strength of the seed, $\Omega_A$, results in larger critical coupling for a fixed $\Omega_B$. This in turns allows the subsystems to be coupled more strongly before entering the unbroken-symmetry phase.

\begin{figure}
    \centering
    \includegraphics[width=\columnwidth]{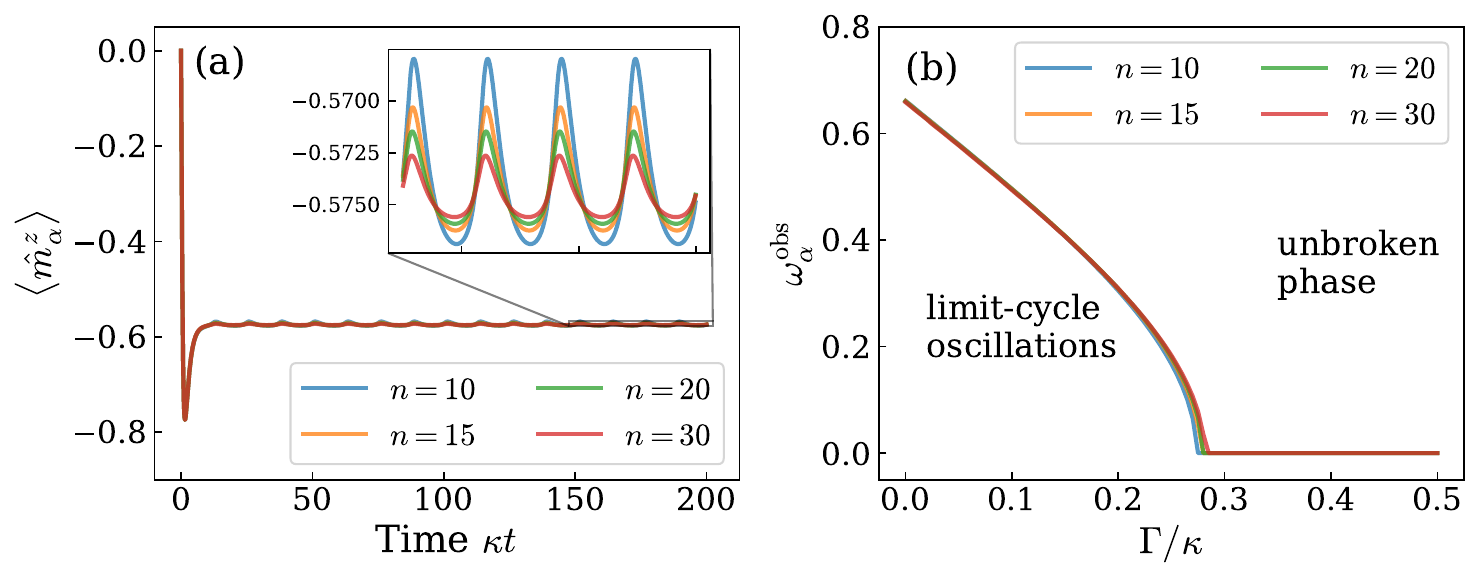}
    \caption{(a) Seeding of crystallization in time for a single seed with $\Omega_{\text{seed}}/\kappa = 1.2$, and $n-1$ subsystems set in the unbroken-symmetry phase, $\Omega_{\alpha}/\kappa=0.9$ for all $\alpha$, with coupling strength $\Gamma/\kappa=0.1$. Seeding can be observed for all subsystems $\alpha$ with identical limit-cycle oscillations of $\langle\hat{m}^z_{\alpha}\rangle$. (b) Observed frequency $\omega^{\text{obs}}_{\alpha}$ vanishes beyond a critical coupling signifying all subsystems entering the unbroken phase.}
    \label{fig:figure2}
\end{figure}

In Fig.~\ref{fig:figure1}(f), we show scaling of the dominant eigenvalue of the Liouvillian spectrum for increasing subsystem size.
For sub-critical coupling, $\text{Re}(\lambda)$ converges to zero while $\text{Im}(\lambda)$ saturates to a finite value predicted by the mean-field analysis indicated by a black arrow.
For critical coupling, $\lambda$ converges to a large negative component indicating absence of a broken-symmetry phase.
The asymptotic eigenvalues were determined by fitting numerical eigenvalues to a large-$N_{\alpha}$ expansion with free coefficients, as shown in Sec.~C of the Appendix.
We also verified that in the case of oscillating coherences the next dominant eigenvalue indeed saturates to a non-zero real value with a large gap (see Appendix \ref{sec:appendix-C}).
To further corroborate our results, we plot time-dependent oscillations for up to $N_{A/B}=100$ which oscillate with frequency predicted by the mean-field, as shown in Fig.~\ref{fig:figure1}(c)-(d).

We now expand our discussion of seeding to $n$ ensembles.
We consider a single seed and $n-1$ subsystems in the unbroken-symmetry phase.
Fig.~\ref{fig:figure2}(a) shows that seeding is observed even for large number of subsystems.
The amplitude of limit-cycle oscillations decreases with increasing $n$ as seen in the inset of Fig.~\ref{fig:figure2}(a).
Increasing the coupling strength $\Gamma/\kappa$ has the effect of boosting this amplitude, however beyond a critical coupling the dissipation becomes unsustainable and limit-cycle oscillations vanish.
This can be seen in Fig.~\ref{fig:figure2}(b) where we plot the \emph{observed frequency} defined as $\omega^{\text{obs}}_{\alpha} = 2\pi / T_{\alpha}$, where $T_{\alpha}$ is the period of oscillation for subsystem $\alpha$.
The observed frequencies of the $n-1$ subsystems follow the frequency of the seed which decreases with increasing coupling strength $\Gamma$ and vanishes upon reaching the critical coupling strength $\Gamma_{\text{crit}}$.
Note that this critical coupling depends on the number of subsystems $n$.

Until this point, we have discussed dissipative coupling of Eq.~(\ref{eq:master_eq}) only.
Seeding of crystallization is observed for coherent exchange interaction of the form $\hat{S}^+_{\alpha}\hat{S}^-_{\beta} + \hat{S}^-_{\alpha}\hat{S}^+_{\beta}$.
The qualitative features of seeding for this type of coupling are the same as for dissipative coupling described above (see Appendix \ref{sec:appendix-D}).

\textit{Synchronization.---}
We now consider the dynamics of Eq.~(\ref{eq:master_eq}) with $n=2$, where both subsystems are in the broken-symmetry phase displaying limit-cycle oscillations with different observed frequencies, $\omega^{\text{obs}}_A \neq \omega^{\text{obs}}_B$, achieved by setting $\Omega_A \neq \Omega_B$.
In the rest of this section, $\Omega_A - \Omega_B$ serves as a convenient measure of detuning between the subsystems.

We are interested in the dynamics of the coupled system, in particular how the observed frequency $\omega^{\text{obs}}_{\alpha}$ changes with varying coupling strength and detuning in anticipation of observing synchronization.
This requires a suitable quantifier of how synchronized the two subsystems are.
Recent years have seen a substantial effort to develop tools to quantify synchronization in quantum systems \cite{jaseem2020generalized,galve2017book}.
An intuitive measure of synchronization for such mean field analysis is the magnitude of the difference between observed frequencies of the two subsystems,
\begin{equation}
    \Delta^{\text{obs}} = \left| \omega_A^{\text{obs}} - \omega_B^{\text{obs}} \right|.
    \label{eq:Delta_obs}
\end{equation}
It is straightforward to see that if the limit-cycle oscillations of both subsystems are synchronized then $\Delta^{\text{obs}}$ vanishes and is positive otherwise.

\begin{figure}[t]
    \centering
    \includegraphics[width=\columnwidth]{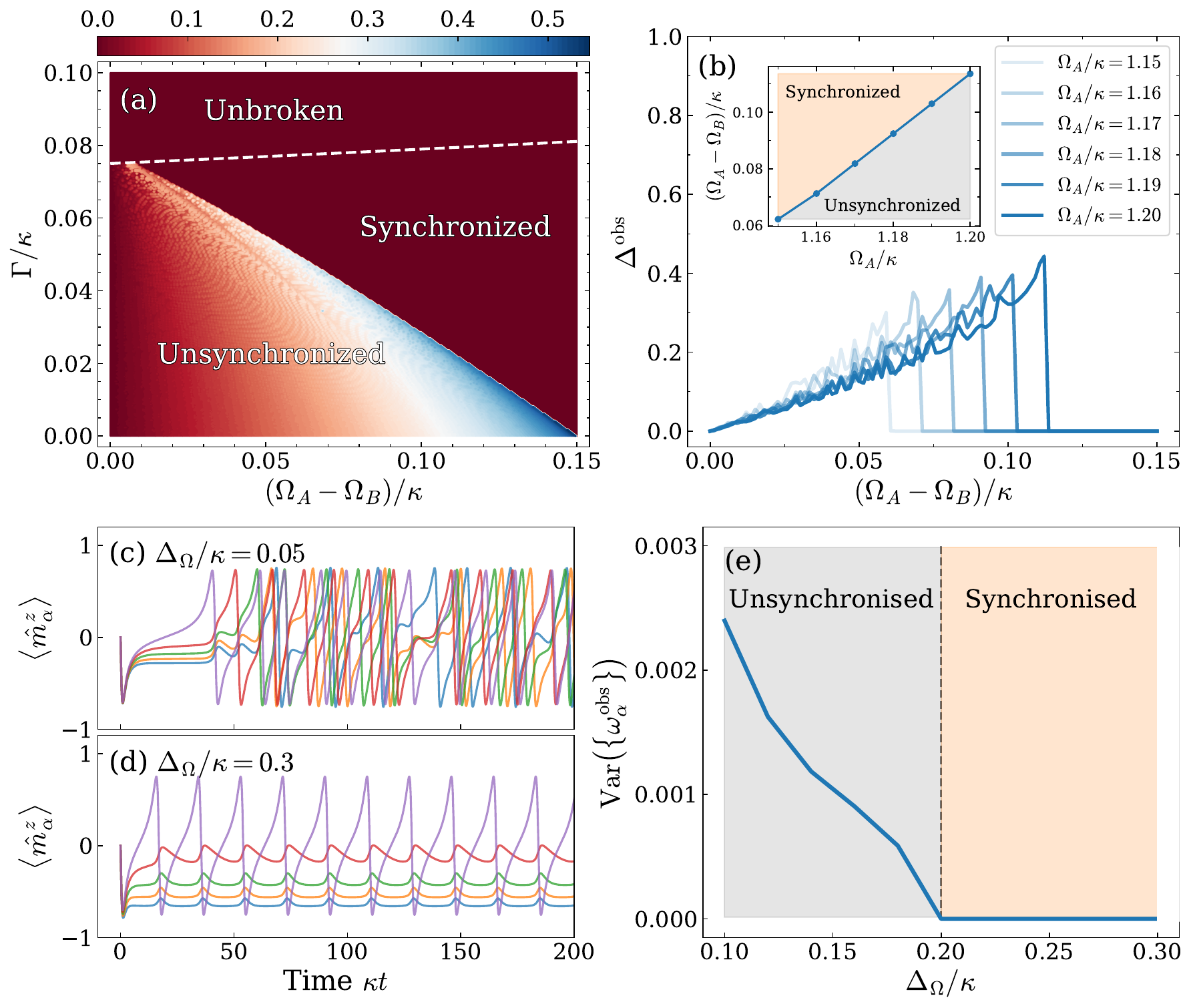}
    \caption{(a) $\Delta^{\text{obs}}$ indicating the synchronization region for $n=2$ and $\Omega_A/\kappa=1.15$. The white dashed curve is the critical coupling from Eq.~(\ref{eq:Gamma_crit}). (b) Increasing $\Omega_A/\kappa$ also increases the minimum detuning required in order to observe synchronization. The coupling strength is kept at $\Gamma/\kappa=0.05$. (c)-(d) Time evolution of $\langle\hat{m}^z_{\alpha}\rangle$ for detuning range $\Delta_{\Omega}/\kappa=0.05$ and $0.3$, respectively. (e) Variance of the observed frequencies $\text{Var}(\{\omega_{\alpha}^{\text{obs}}\})$ vanishes when the detuning interval  $\Delta_{\Omega}/\kappa$ is high enough, indicating synchronized dynamics. Parameters for (c)-(e) are $n=5$, $\max\{\Omega_{\alpha}/\kappa\}=1.5$, and $\Gamma/\kappa=0.5$.}
    \label{fig:figure3}
\end{figure}

Fig.~\ref{fig:figure3}(a) shows $\Delta^{\text{obs}}$ as a function of the detuning $(\Omega_A-\Omega_B)/\kappa$ and the coupling strength $\Gamma/\kappa$.
The driving strength of subsystem $A$ is fixed to $\Omega_A/\kappa=1.15$ and we vary $\Omega_B$.
For reference, we also plot the critical driving strength $\Gamma_{\text{crit}}/\kappa$ of Eq.~(\ref{eq:Gamma_crit}) beyond which both subsystems enter the strongly dissipative regime.

 In Fig.~\ref{fig:figure3}(a) we display the synchronisation measure as a function of detuning and coupling strength.
 We observe that larger detuning, given by large $(\Omega_A-\Omega_B)/\kappa$, requires weak coupling strength $\Gamma/\kappa$ to produce synchronization of the two subsystems.
On the other hand, stronger coupling must be applied when the two subsystems are weakly detuned in order to synchronize them. We believe this is the first time such behaviour has been observed.
This is in contrast with countless examples of classical and quantum synchronizing dynamics studied in literature which share a common property, namely that small detuning requires weak coupling in order to produce synchronization while large detuning requires stronger coupling \cite{pikovsky2003synchronization,sonar2018squeezing,kato2019semiclassical,cabot2021metastable,jaseem2020quantum}.

We now show that this unusual observation can be explained using seeding of crystallization.
Consider a particular detuning in Fig.~\ref{fig:figure3}(a) where $\Omega_A > \Omega_B$.
For weak coupling $\Gamma/\kappa$ the oscillations of both subsystems are unsynchronized.
Increasing the coupling strength has two effects.
It affects the observed frequency of oscillations of the subsystems on one hand, but also acts as a new source of dissipation countering the effect of coherent drives $\Omega_A$ and $\Omega_B$.
Therefore, increasing the coupling strength eventually forces subsystem $B$ into a regime where the coherent drive $\Omega_B$ is not strong enough to sustain its oscillations.
At this point subsystem $A$ starts seeding oscillations to subsystem $B$, forcing it to lock to its frequency of oscillations as previously seen in Fig.~\ref{fig:figure1}(e).

Seeding of crystallization as the mechanism behind synchronization of CTCs has another consequence.
The shape of the phase diagram in Fig.~\ref{fig:figure3}(a) depends not only on the detuning and the coupling strength, but also on the coherent driving strengths $\Omega_A$ and $\Omega_B$.
To see this, we fix the coupling strength $\Gamma/\kappa$ and increase the coherent drive strength $\Omega_A/\kappa$.
We expect that stronger coherent drive $\Omega_A/\kappa$ has the effect of shifting the boundary between unsynchronized and synchronized dynamics towards larger values of detuning $(\Omega_A-\Omega_B)/\kappa$.
In Fig.~\ref{fig:figure3}(b) we show $\Delta^{\text{obs}}$ as a function of the detuning between the CTCs for increasing values of $\Omega_A/\kappa$.
We observe that increasing $\Omega_A/\kappa$ for a fixed coupling strength $\Gamma/\kappa$ produces larger minimum detuning $(\Omega_A-\Omega_B)/\kappa$ required for synchronization as depicted in the inset of Fig.~\ref{fig:figure3}(b).
This confirms our expectation.

We now turn to the case of multiple subsystems, each with its own coherent drive strength $\Omega_{\alpha}$ picked from a detuning range $\Delta_{\Omega}$.
Without loss of generality we assume that the set $\{\Omega_{\alpha}\}$ are distributed uniformly within $\Delta_{\Omega}$ in steps of $\Delta_{\Omega}/(n-1)$.
Fig.~\ref{fig:figure3}(c)-(d) display the time evolution of $\langle\hat{m}^z_{\alpha}\rangle$ for $n=5$ subsystems.
When the detuning interval $\Delta_{\Omega}$ is small, meaning the subsystems are only weakly detuned, no synchronization is observed as seen in Fig.~\ref{fig:figure3}(c).
Increasing $\Delta_{\Omega}$ beyond a critical value produces synchronization in the whole system as seen in Fig.~\ref{fig:figure3}(d).
Generalization of the synchronization measure in Eq.~(\ref{eq:Delta_obs}) is given by the variance of the observed frequencies, $\text{Var}(\{\omega_{\alpha}^{\text{obs}}\})$, which vanishes when the subsystems become synchronized and is finite otherwise.
Fig.~\ref{fig:figure3}(e) shows $\text{Var}(\{\omega_{\alpha}^{\text{obs}}\})$ for fixed coupling strength $\Gamma/\kappa$ and increasing detuning intervals $\Delta_{\Omega}$ indicating clearly the transition from unsynchronized dynamics when the sybsystems are weakly detuned, to synchronized dynamics for large detunings.

Such counter-intuitive behavior is not observed when the subsystems are coupled coherently.
In this case, slightly detuned subsystems require weak coupling to synchronize their dynamics while large detuning demands strong coupling, conforming to the usual behavior in synchronization theory.

\textit{Conclusions.---}
Crystal structure and its formation is one of the fundamental aspects of understanding the solid state. The ability for nucleation centers to seed crystallization in a solute and its role in spontaneous spatial symmetry breaking has been appreciated for a long time.
Results presented in this manuscript contribute towards understanding the foundations of spontaneous continuous time-translation symmetry breaking in coupled systems, an area which has gone underexplored so far.
Seeding of crystallization in time is not only possible but in many cases inevitable as it occurs for both coherent as well as dissipative coupling and a broad range of parameters.
The role of dissipation in spontaneous time-translation breaking is to destroy persistent limit-cycle oscillations.
It is therefore surprising that dissipative coupling leads to proliferation of broken symmetry phases through the global system.
Given appropriate coupling, we expect that other physical systems, whether displaying continuous or discrete time-translational symmetry breaking, can be used to observe such behavior.

We note that the emergence of oscillations in networks of otherwise quiescent classical oscillators has been investigated in \cite{biswas2019oscillatory}.
Unlike the seeding effect discussed in this manuscript, the oscillations either spread only to a fraction of the subsystems or are observed only during the transient dynamics.
Results presented in this manuscript are among the first steps toward understanding similar phenomena in quantum systems.

In the context of mutual synchronization, seeding leads to the extraordinary observation that dissipatively coupled CTCs do not follow the principle tennet of synchronization theory; that increasing detuning requires stronger coupling in order for the subsystems to synchronize.
In fact, the precise opposite is true in our case.
Conventional and nature-inspired synchronization schemes have numerous practical applications in complex networks \cite{choi2017principles}.
Researchers and engineers go to great lengths designing dynamical systems which are nearly identical in order to facilitate their synchronization in these protocols.
Our results show that this is not necessarily needed, opening the path to new design philosophies in network synchronization.

% \begin{acknowledgments}
\textit{Acknowledgments.---} The authors would like to acknowledge useful conversations with Punit Parmananda. Numerics were performed using QuTiP \cite{johansson2012qutip,johansson2013qutip2} and Python library PIQS \cite{shammah2018open}. M.H. is supported by MEXT Quantum Leap Flagship Program Grants No. JPMXS0118067285 and No. JPMXS0120319794. S.V. acknowledges support from a DST-SERB Early Career Research Award (ECR/2018/000957) and DST-QUEST grant number DST/ICPS/QuST/Theme-4/2019.
% \end{acknowledgments}

% \bibliographystyle{apsrev4-1}
% \bibliography{references}


\appendix
\onecolumngrid

\section{Master equation}
\label{sec:appendix-A}

In this section, we provide a brief derivation of the master equation used in Eq.~(3) of the main text. Consider $n$ ensembles, each consisting of identical atomic dipoles with total ensemble spin $S_{\alpha}$.
We denote the dipole annihilation operator by $\hat{\sigma}^-_{j,\alpha}$, where $j$ labels the atomic dipole within ensemble $\alpha$.
Each dipole within an ensemble interacts with a bosonic field with annihilation operator $\hat{a}_{\alpha}$ via energy-conserving exchange Hamiltonian with strength $g_{\alpha}$.
The independent bosonic fields are also coherently driven on resonance with strength $\eta_{\alpha}$.
Finally, all ensembles interact with another independent bosonic field $\hat{c}$ with strength $g_c$.
All bosonic field are damped a rates given by $\gamma_{\alpha}$ and $\gamma_c$.

Total Hamiltonian in the frame rotating at the driving frequency is then
\begin{align}
    \hat{H} & = \sum_{j, \alpha} \left\{ \frac{g_{\alpha}}{\sqrt{S_{\alpha}}} \left( \hat{a}^{\dagger}_{\alpha} \hat{\sigma}^-_{j,\alpha} + \hat{\sigma}^+_{j,\alpha}\hat{a}_{\alpha} \right) -i\eta_{\alpha} \left( \hat{a}_{\alpha} - \hat{a}_{\alpha}^{\dagger} \right) + \frac{g_c}{\sqrt{S}} \left( \hat{c}^{\dagger} \hat{\sigma}^-_{j,\alpha} + \hat{\sigma}^+_{j,\alpha}\hat{c} \right) \right\}.
\end{align}
Dynamics of the total system is described by the following master equation,
\begin{equation}
    \dot{\rho} = -i \left[\hat{H}, \rho \right] + \sum_{\alpha}\gamma_{\alpha}\mathcal{D}[\hat{a}_{\alpha}]\rho + \gamma_C\mathcal{D}[\hat{c}]\rho.
\end{equation}
Equations of motion for the bosonic degrees of freedom are
\begin{equation}
    \frac{d\hat{a}_{\alpha}}{dt} = -\frac{ig_{\alpha}}{\sqrt{S_{\alpha}}}\hat{S}_{\alpha}^- + \eta_{\alpha} - \frac{\gamma_{\alpha}}{2}\hat{a}_{\alpha}, \qquad
    \frac{d\hat{c}}{dt} = -\frac{ig_c}{\sqrt{S}}\hat{S}^- - \frac{\gamma_c}{2}\hat{c},
\end{equation}
where $\hat{S}_{\alpha}^- = \sum_{j} \hat{\sigma}_{j,\alpha}^-$, $\hat{S}^- = \sum_{\alpha} \hat{S}_{\alpha}^-$, and $S = \sum_{\alpha} S_{\alpha}$.

In the bad cavity limit, we can set the time derivatives to zero,
\begin{equation}
    \hat{a}_{\alpha} = -\frac{2ig_{\alpha}}{\gamma_{\alpha}\sqrt{S_{\alpha}}} \hat{S}_{\alpha}^- + \frac{2\eta_{\alpha}}{\gamma_{\alpha}}, \qquad \hat{c} = -\frac{2ig_c}{\gamma_c\sqrt{S}} \hat{S}^-.
\end{equation}
Substituting back into the master equation,
\begin{align}
    \dot{\rho} = -i \left[ \sum_{\alpha} \frac{4 g_{\alpha}\eta_{\alpha}}{\gamma_{\alpha}\sqrt{S_{\alpha}}} \hat{S}_{\alpha}^x, \rho \right] + \sum_{\alpha} \frac{4g_{\alpha}^2}{\gamma_{\alpha} S_{\alpha}} \mathcal{D}[\hat{S}_{\alpha}^-] \rho + \frac{4g_c^2}{\gamma_c S}\mathcal{D}[\hat{S}^-]\rho.
\end{align}
Defining
\begin{equation}
    \Omega_{\alpha} \equiv \frac{2g\eta_{\alpha}}{\gamma\sqrt{S_{\alpha}}}, \qquad \kappa_{\alpha} \equiv \frac{4g_{\alpha}^2}{\gamma_{\alpha} S_{\alpha}}, \qquad \Gamma \equiv \frac{4g_c^2}{\gamma_c S},
\end{equation}
we recover the master equation in the main text.

\section{Thermodynamic limit}
\label{sec:appendix-B}

Using Eq.~(3) we can write a system of coupled nonlinear differential equations for the expectation values of the rescaled spin operators $\hat{m}_{\alpha}^k \equiv \langle\hat{S}_{\alpha}^k\rangle/S_{\alpha}$ when $S_{\alpha}\rightarrow\infty$ for all subsystems,
\begin{align}
	\frac{d\langle\hat{m}^x_{\alpha}\rangle}{dt} & =  \left( \kappa + \frac{\Gamma}{n} \right) \langle\hat{m}^x_{\alpha}\rangle \langle\hat{m}^z_{\alpha}\rangle + \frac{\Gamma}{n} \langle\hat{m}^z_{\alpha}\rangle \sum_{\beta\neq\alpha} \langle\hat{m}^x_{\beta}\rangle, \nonumber\\
	\frac{d\langle\hat{m}^y_{\alpha}\rangle}{dt} & =  -\Omega_{\alpha} \langle\hat{m}^z_{\alpha}\rangle + \left( \kappa + \frac{\Gamma}{n} \right) \langle\hat{m}^y_{\alpha}\rangle \langle\hat{m}^z_{\alpha}\rangle + \frac{\Gamma}{n} \langle\hat{m}^z_{\alpha}\rangle \sum_{\beta\neq\alpha} \langle\hat{m}^y_{\beta}\rangle, \\
	\frac{d\langle\hat{m}^z_{\alpha}\rangle}{dt} & =  \Omega_{\alpha} \langle\hat{m}^y_{\alpha}\rangle - \left( \kappa + \frac{\Gamma}{n} \right) \left[ \langle\hat{m}^x_{\alpha}\rangle^2 + \langle\hat{m}^y_{\alpha}\rangle^2 \right] - \frac{\Gamma}{n} \left[ \langle\hat{m}^x_{\alpha}\rangle \sum_{\beta\neq\alpha} \langle\hat{m}^x_{\beta}\rangle + \langle\hat{m}^y_{\alpha}\rangle \sum_{\beta\neq\alpha} \langle\hat{m}^y_{\beta}\rangle \right]. \nonumber
\end{align}

To get an intuition about the dynamics of the above nonlinear system of differential equations, we consider the simple case of $n=2$ oscillators, labelled $A$ and $B$.
We begin by finding a fixed point of the system by setting the time derivatives to 0.
The fixed point has the following form,
\begin{equation}
    \{m^{x^*}_{A},m^{y^*}_A,m^{z^*}_A,m^{x^*}_{B},m^{y^*}_B,m^{z^*}_B \}=\{0,\frac{\frac{\Gamma}{2}(\Omega_A-\Omega_B)+\kappa\Omega_A}{\kappa(\Gamma+\kappa)},0,0,\frac{\frac{\Gamma}{2}(\Omega_B-\Omega_A)+\kappa\Omega_B}{\kappa(\Gamma+\kappa)},0 \}.\label{eq:fixed_point}
\end{equation}
The eigenvalues of the Jacobian matrix at the fixed point either vanish or are purely real, suggesting Eq.~(\ref{eq:fixed_point}) is a hyperbolic fixed point.
The magnitude of the rescaled spin is normalized to unity, meaning the fixed point is physical only if $m^{y^*}_{\alpha} \leq 1$.
This leads to a constraint on the coupling strength,
\begin{equation}
    \Gamma \geq \frac{2\kappa(\Omega_A - \kappa)}{2\kappa - (\Omega_A - \Omega_B)} \equiv \Gamma_{\text{crit}}^1, \qquad\text{and}\qquad
    \Gamma \geq \frac{2\kappa(\Omega_B - \kappa)}{2\kappa - (\Omega_B - \Omega_A)} \equiv \Gamma_{\text{crit}}^2.
\end{equation}
The hyperbolic fixed point is a physical state when $\Gamma \geq \Gamma_{_\text{crit}} \equiv \max\{\Gamma_{\text{crit}}^1, \Gamma_{\text{crit}}^2\}$ and corresponds to the unbroken symmetry phase of the system.
As soon as $\Gamma < \Gamma_{\text{crit}}$, both subsystems develop limit-cycle oscillations and both break the continuous time-translation symmetry.
Without loss of generality, we focus on the case when $\Omega_A \geq \Omega_B$ and we have $\Gamma_{\text{crit}} = \Gamma_{\text{crit}}^1$ which is Eq.~(4) of main text.

\section{Large-$N_{\alpha}$ expansion of the eigenvalues}
\label{sec:appendix-C}

We now discuss the spectrum of the Liouville superoperator $\mathcal{L}$ generating the dynamics of Eq.~(3) for the case of $n=2$.
We are interested in the functional dependence of the spectrum on the increasing size of the subsystems $\alpha$, on the system parameters and how these ultimately affect the values to which the eigenvalues converge as $N_{\alpha}=2S_{\alpha}\rightarrow\infty$.
The total number of atoms is $N=N_A+N_B=2N_A$.

\begin{figure*}[htp!]
    \centering
    \includegraphics[width=0.8\textwidth]{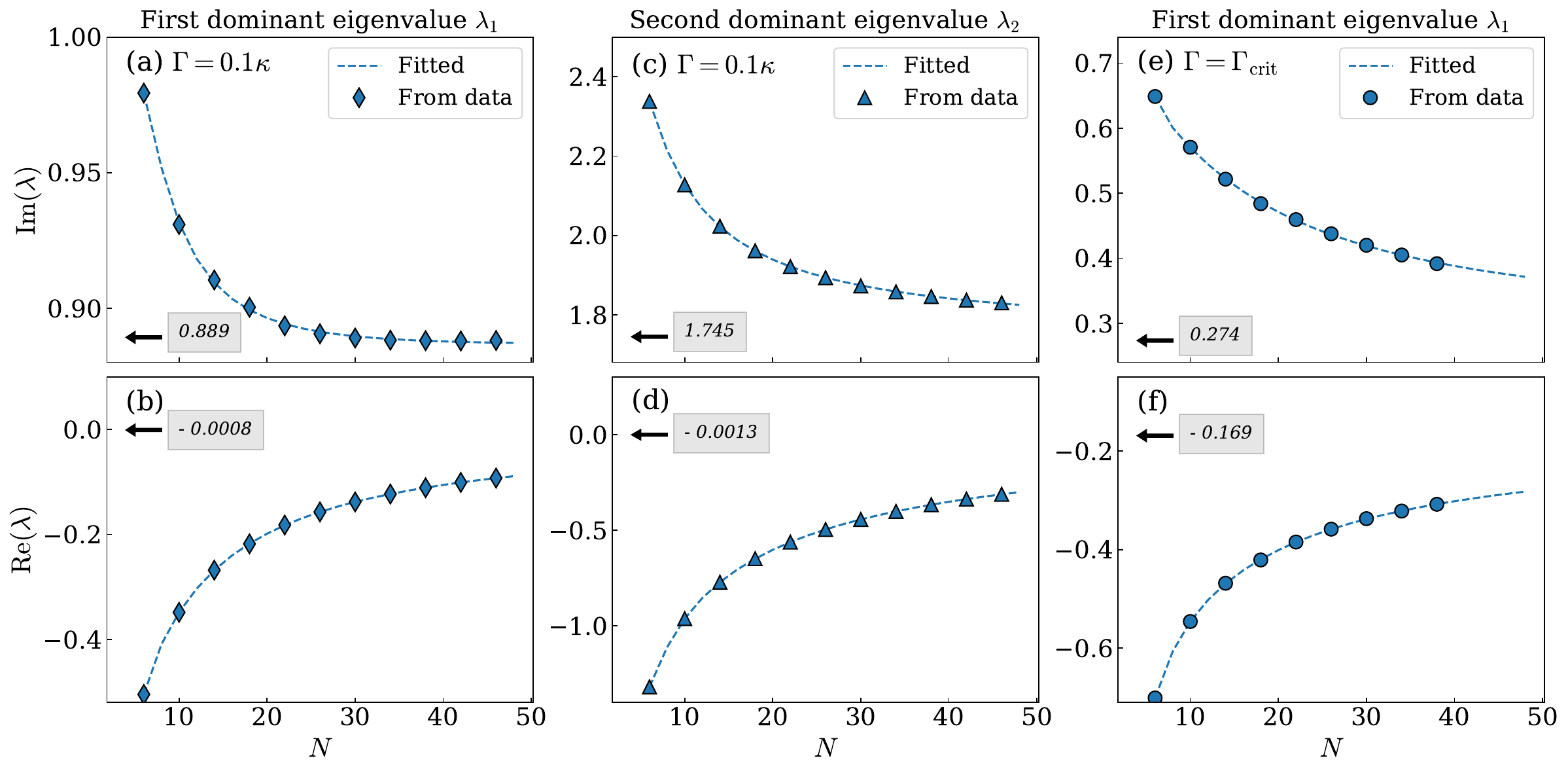}
    \caption{Functional fitting of eigenvalues of Liouville superoperator corresponding to Eq.~(3) for various values of $N$ with parameters values $\Omega_A=1.5$, $\Omega_B=0.9$ and different values of $\Gamma$. Plots (a-b) and (c-d) show functional fitting of real and imaginary parts of the first and second dominant eigenvalues respectively for $\Gamma=0.1$. Plots (e-f) represents the first dominant eigenvalue for $\Gamma=\Gamma_{\text{crit}}$. The text in gray box following the arrow in each subplot represents the corresponding value of real and imaginary part of eigenvalues for $N\rightarrow \infty$.}
    \label{fig:L_eigs}
\end{figure*}

Real and imaginary parts of the eigenvalues are fitted using function $F(N)=\sum_{i=0}^{\mu} a_i/N^i$ where $\mu=5$ for real and $\mu=4$ for imaginary parts to get best fit as shown in Fig.~\ref{fig:L_eigs}.
As $N \rightarrow \infty$, $F(\infty)\rightarrow a_0$ gives the value of corresponding real or imaginary part in thermodynamic limit.
Fig.~\ref{fig:L_eigs} depicts that only the first dominant eigenvalue $\lambda_1$ has a vanishing real part in the thermodynamic limit, see Fig.~\ref{fig:L_eigs}(a),  for $\Gamma=0.1<\Gamma_{\text{crit}}$  and this suggests that non-equilibrium state of the system will be oscillating with frequency equal to corresponding imaginary part shown in Fig.~\ref{fig:L_eigs}(b).
This result is in agreement with the frequency spectrum shown by mean-field analysis in Fig.~1(e) of the main text.
The second dominant eigenvalue $\lambda_2$ for $\Gamma=0.1$ is shown in Fig.~\ref{fig:L_eigs}(c)-(d) and we observe a small but real part relatively more
negative than $\text{Re}(\lambda_1)$, which results in dissipation of oscillations in longer time limit.
As discussed previously, for $\Gamma\geq \Gamma_{\text{crit}}$ the entire system settles to an unbroken-symmetry phase as there are no pure imaginary eigenvalues of $\mathcal{L}$.
This can be observed in Fig.~\ref{fig:L_eigs}(e)-(f) where the real part of the first dominant eigenvalue is non-zero in the thermodynamic limit and hence no broken-symmetry phase exists for $\Gamma=\Gamma_{\text{crit}}$.

\section{Seeding for coherent coupling.}
\label{sec:appendix-D}

We now consider coherent coupling between the subsystems and show that seeding is observed.
The master equation is the following,
\begin{equation}
    \dot{\rho} = -i \sum_{\alpha} \left\{ \left[ \Omega_{\alpha} S^x_{\alpha} + \frac{\Gamma}{2S} \sum_{\beta\neq\alpha} \left( S^+_{\alpha} S^-_{\beta} + S^-_{\alpha} S^+_{\beta} \right), \rho \right] + \frac{\kappa}{S_{\alpha}} \mathcal{D}[S^-_{\alpha}]\rho \right\}.
\end{equation}
The corresponding classical system of dynamical equations for the rescaled spin operators and diverging $S_{\alpha}$ is
\begin{align}
    \frac{d\langle\hat{m}^x_{\alpha}\rangle}{dt} & = \kappa \langle\hat{m}^x_{\alpha}\rangle \langle\hat{m}^z_{\alpha}\rangle + \frac{\Gamma}{n} \langle\hat{m}^z_{\alpha}\rangle \sum_{\beta\neq\alpha} \langle\hat{m}^y_{\beta}\rangle, \nonumber\\
    \frac{d\langle\hat{m}^y_{\alpha}\rangle}{dt} & = -\Omega_{\alpha} \langle\hat{m}^z_{\alpha}\rangle + \kappa \langle\hat{m}^y_{\alpha}\rangle \langle\hat{m}^z_{\alpha}\rangle - \frac{\Gamma}{n} \langle\hat{m}^z_{\alpha}\rangle \sum_{\beta\neq\alpha} \langle\hat{m}^x_{\beta}\rangle, \\
    \frac{d\langle\hat{m}^z_{\alpha}\rangle}{dt} & = \Omega_{\alpha} \langle\hat{m}^y_{\alpha}\rangle - \kappa \left[ \langle\hat{m}^x_{\alpha}\rangle^2 + \langle\hat{m}^y_{\alpha}\rangle^2 \right] + \frac{\Gamma}{n} \left[ \langle\hat{m}^y_{\alpha}\rangle \sum_{\beta\neq\alpha} \langle\hat{m}^x_{\beta}\rangle - \langle\hat{m}^x_{\alpha}\rangle \sum_{\beta\neq\alpha} \langle\hat{m}^y_{\beta}\rangle \right]. \nonumber
\end{align}

\begin{figure*}[htp!]
    \centering
    \includegraphics[width=0.9\textwidth]{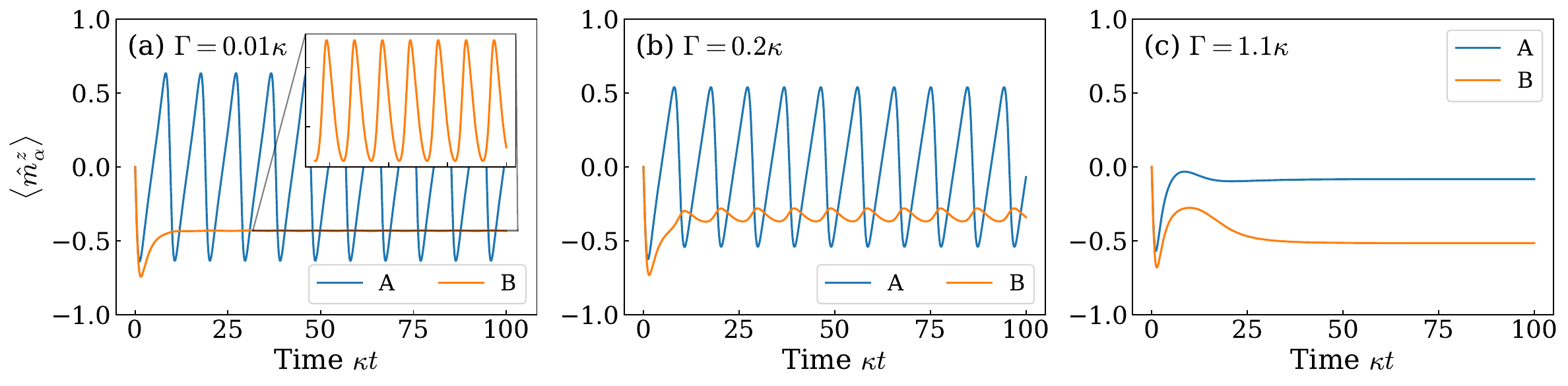}
    \caption{Evolution of $\langle\hat{m}^z_{\alpha}\rangle$ for increasing values of coupling strength. Subsystem A is the seed with $\Omega_{\text{seed}}/\kappa=1.2$ while $\Omega_B/\kappa = 0.9$. Seeding is observed in (a)-(b) while in (c) the coupling strength $\Gamma>\Gamma_{\text{crit}}$ and no limit-cycle oscillations persist.}
    \label{fig:figure5_appendix}
\end{figure*}

We consider the case of two CTCs, $n=2$, in the large-local spin limit $S_{\alpha}\rightarrow\infty$.
Subsystem $A$ is the seed with $\Omega_{\text{seed}}/\kappa = 1.2$ while subsystem $B$ is in the unbroken phase with $\Omega_B/\kappa=0.9$.
Fig.~\ref{fig:figure5_appendix}(a) shows that even for very weak coherent coupling, subsystem $B$ develops persistent limit-cycle oscillations, breaking the time-translation symmetry.
The amplitude of these oscillations increases with the coupling strength as seen in Fig.~\ref{fig:figure5_appendix}(b).
However, similar to the case of dissipative interaction, when the coupling becomes too strong both subsystems lose thier limit-cycle oscillations and return to the unbroken-symmetry phase.
The critical coupling strength in the case of coherent coupling was determined numerically to be $1.0882 < \Gamma_{\text{crit}}/\kappa < 1.0883$.

\end{document}